\documentclass[12pt]{article}

\usepackage[paper=a4paper, margin=2cm]{geometry}
\usepackage[russian,english]{babel}
\usepackage[cp1251]{inputenc}
\usepackage{amsmath}

\usepackage[logos]{amsbib}

\begin{document}

\date{}
\renewcommand{\refname}{References}

\author{A.~Yu.~Samarin
}
\title{Quantum evolution in terms of mechanical motion}

\maketitle

\centerline{Samara Technical State University, 443100 Samara, Russia}
\centerline{}

\abstract{Quantum tunneling is considered from the point of view of local realism. It is concluded that a quantum object tunneling through a potential barrier cannot be interpreted as a point-like particle because such an interpretation generates a contradiction with the impossibility of faster-than-light motion. Such a contradiction does not arise if a quantum object is considered as a continuous medium formed by the fields of matter. The dynamics law of the mechanical motion of these matter fields is derived from the quantum evolution law in the path integrals form. The analysis of tunneling shows that this dynamics law has a form of the principle of least action with a complex time variable. The approach used here is not only a physical interpretation of quantum tunneling consistent with special relativity but is also applicable to the description of a wide range of quantum phenomena for which traditional research methods are impracticable.} 



\section{Introduction}
The phrase "the tunneling of a quantum particle" contains the interpretation of a quantum object as a point-like particle. To avoid ambiguity, instead of the term "quantum particle" the term "quantum object" will be used to mean something material and homogeneous described by the wave function depending on the radius vector in physical space, and the term "corpuscle"~--- a point-like particle. The fact that a local external effect on a quantum object instantly changes its wave function throughout space creates difficulties in the interpretation of the wave function as a characteristic of a corpuscle in terms of local realism~\cite{bib:1} (all phenomena are considered here from  the point of view of local realism). This impossibility becomes apparent when considering the phenomena associated with the reduction of the wave function~\cite{bib:2}. Acceptance of the corpuscle concept, when describing the reduction of a multipartite quantum system, causes contradiction between quantum mechanics, considered as a local realistic theory, and special relativity~\cite{bib:1}. This contradiction has been verified experimentally~\cite{bib:3,bib:4,bib:5,bib:6,bib:7} in accordance with the scheme proposed in~\cite{bib:81}, and a lot of efforts have been made to eliminate it, e.g.~\cite{bib:8,bib:9,bib:10}. All of them are based on the assertion that, since the result of the collapse is probabilistic, then the transfer of information using this process is impossible. Then, the term "information transfer" is given the meaning of signal transmission, and, based on this, it is concluded that there is no faster-than-light motion in the reduction process. The question of the non-local nature of causal relationship between the effect on one particle and the instantaneous change of the state of another one, remote in space, remains unanswered. However, the questionable rigor of such deductive reasoning is not the subject of this study.

The concept of a corpuscle is clearly unacceptable when considering tunneling. In accordance with~\cite{bib:11} there is no appreciable delay in the transmission of a wave packet through a barrier. The transmission time does not depend on the barrier thickness (the Hartman effect)~\cite{bib:12}. Thus, if we consider a wave packet as a mathematical object describing possible positions of the corpuscle, then the speed of this corpuscle can exceed the speed of light. The experimental studies~\cite{bib:13,bib:14} report that the ionization time of the hydrogen atom in the tunneling process is close to zero. The paper~\cite{bib:14} puts an upper bound on tunnelling time, which is less than the value of any time which can theoretically be considered as the  tunnelling time of a corpuscle~\cite{bib:15,bib:16}. This is consistent with the theoretical findings~\cite{bib:11,bib:12} and adds confidence that tunneling time is zero. There is disagreement about the definition of traversal time. The quantum mechanics postulate corresponding to the Borne rule~\cite{bib:17} determines the probability of the measurement result~\cite{bib:2}. That is, with respect to the measurement of the position of a quantum object, it gives the probability of the detector triggering in a small volume in vicinity of a certain point in space, but does not give the probability of finding a corpuscle in this volume. Orthodox quantum mechanics, the mathematical structure of which does not contain the concept of a corpuscle, forces us to define this term as time delay between equal values of the phases of stationary wave functions on opposite sides of the barrier. If the total energy is less than the potential energy in the barrier region, then the stationary wave functions do not have a phase factor depending on space coordinates and the phase relationships between them are the same throughout the barrier region. Thus, tunneling time defined in this way is zero. This fact does not contradict special relativity until we interpret a quantum object as a corpuscle. If such an object is a corpuscle, then, taking into account the Born rule and the statistical definition of probability, we are forced to admit that some of the particles falling on the barrier from one side instantly shift on the other one. This is a direct inadmissible contradiction with special relativity which indicates that the concept of a point-like particle cannot be used to interpret quantum mechanics phenomena.

To escape this contradiction, a homogeneous quantum object has to be considered as a continuous medium. The possibility of its identification with the wave function is doubtful. This is, due to the property of a substance, to change its position in space exclusively as a result of motion (this property is mathematically expressed by the continuity equation). The wave function does not have such a property (this is obvious when the wave function collapses). In addition, since vacuum has physical characteristics, matter is also present where wave functions of all material objects are zero. 

The properties of this continuous medium should determine all the properties of the wave function considered as its physical characteristic~\cite{bib:82,bib:83}. Then it seems logical to imagine the material support of the wave function in the form of a continuous medium the mechanical motion of which should generate the orthodox dynamics of the wave function. This was done in the articles~\cite{bib:18,bib:19} where both unitary dynamics and the dynamics of the collapse of quantum states were obtained from the mechanical motion of a peculiar continuous medium. However, it is simpler and clearer to do the opposite, namely, to extract the properties of the material support of the wave function from the orthodox law of quantum evolution.

\section{Real paths}
This procedure can be directly realized for the one-dimensional integral wave equation of non-relativistic quantum mechanics in the form~\cite{bib:20,bib:21}:
\begin{equation}\label{eq:math:ex1}
\Psi_t(x)=\int\limits_{-\infty}^\infty K_{t,t_0}(x,x_0)\Psi_{t_0}(x_0)\,dx_0,
\end{equation}
where the kernel of integral evolution operator (transition amplitude) $K_{t,t_0}(x,x_0)$ is determined by the path integral:
\begin{equation}\label{eq:math:ex2}
K_{t,t_0}(x,x_0)=\int \exp{\Bigl(\frac{i}{\hbar}S[x(\tau)]\Bigr)}\,[dx(\tau)],
\end{equation}
All the virtual paths $x(\tau)$ have the same end points $x_0=x(t_0)$ and $x=x(t)$ (the current time variable in paths is hereinafter referred to as the Greek letter $\tau$, while the instants of time corresponding to the end points of paths are referred to as the Latin letter $t$). 

Following the proposed approach, we should assume that the transition amplitude $K_{t,t_0}(x,x_0)$ connects the time dependence of the wave function with the change of the material support position in time. The latter must be uniquely determined by the external and initial conditions. The initial conditions for material points of a continuous medium can be uniquely determined only if this continuous medium is a matter field. In accordance with~\cite{bib:19},  this means that all individual particles must have the same energy and the same direction of velocity. Let a continuous medium be a matter field, then, taking into account the fact that the support is localized in space when $t\rightarrow t_0$ at any time $t_0$, we can derive the law of the mechanical motion of the matter field from the functional equation:
\begin{equation*}
K_{t,t_0}(x,x_0)=F_{t,t_0}[x^m(\tau),x,x_0],
\end{equation*}
where $x^m(t)$ is a unique path defined by this equation; $F_{t,t_0}[x^m(\tau),x,x_0]$ is the sought functional. If we could express the transition amplitude in the form of the functional $F[x^m(t)]$, then the path $x^m(t)$ could be considered as a real path of the material point of the continuous medium. Taking into account the form of the integrand in~\eqref{eq:math:ex2}, the functional $F[x^m(t)]$ is sought in the form
\begin{equation*}
F_{t,t_0}[x^m(\tau),x,x_0]= \exp{\Bigl(\frac{i}{\hbar}S_{t,t_0}^m[x(\tau),x,x_0]\Bigr)}.
\end{equation*}
To find the real path $x^m(\tau)$, the path integral~\eqref{eq:math:ex2} has to be taken. It can be done for the path integral in real form. The quantum path integral can be formally converted into real form by replacing real time $\tau$ with imaginary negative time $-i\tau$~\cite{bib:22} in the expression for the classical action $S$ (The replacement here is only a mathematical procedure and has no physical meaning). In this form, the quantum path integral is similar to the functional integral for the Brownian motion~\cite{bib:23}. But there is a fundamental difference between them. Brownian paths are non-differentiable time functions (due to random collisions), while quantum virtual paths are differentiable (and deterministic). Due to this, the quantum path integral~\eqref{eq:math:ex2} can be represented in the form~\cite{bib:22}

\begin{eqnarray}\label{eq:math:ex3}
K_{t,t_0}(x,x_0)=\lim_{ \varepsilon\to 0}\Biggl(\frac{m}{2\pi\hbar\varepsilon }\Biggr)^{n/2}\nonumber\\
\times \idotsint\exp{\biggl(-\frac{1}{\hbar}S(x_0...x_n,\varepsilon)\biggr)}\prod\limits_{k=1}^{n-1}dx_k,
\end{eqnarray}
where the action $S(x_0...x_n,\varepsilon)$ for $n=(t-t_0)/\varepsilon$ successive infinitesimal time intervals $\varepsilon$ is  
\begin{eqnarray*}
S(x_0...x_n,\varepsilon)\\=\exp{\sum\limits_{k=0}^{k=n-1}\biggl(\frac{m}{2}\Bigl(\frac{x_{k+1}-x_{k}}{\varepsilon}\Bigr)^2+V\Bigl(\frac{x_{k+1}+x_k}{2}\Bigr)\biggr)\varepsilon}.
\end{eqnarray*}
 Thus, the quantum path integral can be represented as the limit of the set of Gaussian integrals (such a limit has no physical meaning for the Brownian functional integral, since, in this case, the time interval $\varepsilon$ cannot be less than the ratio $\lambda/v$, where $\lambda$ is free length, $v$~--- the thermal speed of a Brownian particle). If the linear pieces are small enough to consider $\partial V(x_k)/\partial x_k$ as constants, then each of the Gaussian integrals in~\eqref{eq:math:ex3} can be taken: 
\begin{eqnarray*}
\sqrt{\frac{m}{2\pi\hbar\varepsilon }}\int \exp{\biggl(-\frac{1}{\hbar}S(x_k,x_{k-1})\biggr)}\,dx_k\\=\exp{\biggl(-\frac{1}{\hbar}S(x^m_k,x_{k-1})\biggr)},
\end{eqnarray*}
where $x^m_k$ is the coordinate, defined by the expression
\begin{equation*}
x_k^m=x_{k-1}-\frac{\partial V}{\partial x}\frac{\varepsilon^2}{2m}.
\end{equation*}
On the other hand, this is the coordinate of the maximum of the Gaussian function and, therefore, satisfies the condition
\begin{equation*}
\frac{\partial}{\partial x_k}\exp{\biggl(-\frac{1}{\hbar}S(x_k,x_{k-1})\biggr)}\Biggr|_{x_k=x_k^m}=0
\end{equation*}

As $\varepsilon$ tends to zero, the normalized Gaussian curve of the form
\begin{equation*}
\sqrt{\frac{a}{2\pi}}\exp{\biggl(-\frac{ax_k^2}{2}+bx_k\biggr)},
\end{equation*}
narrows while maintaining the position of the maximum and the area, and in the limit transforms into the Dirac delta function ($a$ and $b$ do not depend on $k$). Thus, we have
\begin{eqnarray*}
\lim_{ \varepsilon\to 0}\sqrt{\frac{m}{2\pi\hbar\varepsilon }}\\\times\int\limits_{-\infty}^\infty \exp\biggl({-\frac{m}{2\hbar\varepsilon}\bigl(x_{k}-x_{k-1}\bigr)^2-\frac{\varepsilon}{\hbar}V\Bigl(\frac{x_{k}+x_{k-1}}{2}\Bigr)}\biggr)\,dx_k\\=\int \exp\bigl({dS(x_k)}\bigr)\delta(x_k-x_k^m)\,dx_k,
\end{eqnarray*}
where
\begin{equation*}
dS(x_k)=\frac{\partial S(x_k,x_{k-1})}{\partial (x_k-x_{k-1})}d(x_k-x_{k-1}).
\end{equation*}
Returning to real time for the transition amplitude~\eqref{eq:math:ex2}, we get
\begin{equation}\label{eq:math:ex4}
K_{t,t_0}(x,x_0)= \delta\bigl(x(t)-x^m(t)\bigr)\exp{\frac{i}{\hbar}S[x^m(\tau)]},
\end{equation}
where the path $x^m(\tau)$ corresponds to the least action. Thus, the path integral~\eqref{eq:math:ex2} in the integral wave equation~\eqref{eq:math:ex1} is just another mathematical representation of the transition amplitude~\eqref{eq:math:ex4} for the path corresponding to the classical motion of a material particle. This path depends on external conditions, the initial position and velocity. Further, we assume that the energies of all individual particles of the continuous medium are the same. This means that a quantum object is in a stationary state, and the continuous medium is a matter field~\cite{bib:19}. The end points of the paths are indicated in the transition amplitude; they uniquely determine the real paths of material particles. For these reasons, the end points of the paths and energy will only be specified if necessary, and, in general, we'll mean $x^m(\tau)\equiv x^m(\tau,x_0,t_0,E)$. The path $x^m(\tau)$ can be found from the condition
\begin{equation}\label{eq:math:ex5}
\delta S[x^m(\tau)]=0.
\end{equation}

The set of Gaussian integrals in~\eqref{eq:math:ex3} does not contain an integral over the position of the end point $x_n$ (the positions of the end points are fixed). The formal result of this is that after integrating the expression~\eqref{eq:math:ex3}, the delta function~$\delta(x_n-x_n^m)$ is preserved. Physically, this means that the transition amplitude of a material particle differs from zero only at that point in space where this particle is currently located.

\section{The mechanical motion law}
The time-independent Schr{\"{o}}dinger equation for the infinite motion has stationary solutions which are the superposition of waves moving in opposite directions for each value of the total energy. It is necessary that the wave function and its spatial derivative should both be continuous. Physically, this means that in the regions where $\partial V/\partial x\neq 0$, the incident wave generates a reflected wave, even if $E>V$.
 Thus, a continuous medium can be formed by one matter field only if the field of potential energy is uniform throughout all space. Otherwise, another matter field is generated, moving in opposite direction. This is a direct consequence of the necessity to satisfy both the continuity equation and the expression for the transition amplitude~\eqref{eq:math:ex4}. Really, if $\partial V/\partial x\neq 0$, then the particle velocity is a coordinate function, and it follows from usual continuity equation that the density distribution of matter should be the corresponding function of spatial coordinates. However, as follows from~\cite{bib:19}, an individual point of the material cannot be considered as an infinitesimal volume of matter and thus has no density at all and the continuity equation cannot be satisfied in this case. The problem of conservation of the substance is solved by itself if we assume the simultaneous existence of two material fields having the same energy and moving in opposite directions. 
 
 Thus, there are simultaneously two opposite paths corresponding to the principle of least action in the region of space, where $\partial V/\partial x\neq 0$. The fundamental difference from classical mechanics is that both of these paths are realised simultaneously. The proportion of the substance of the medium moving in each direction is defined by the principle of conservation of this substance. The dynamics of the motion of the "reflected" matter field can be provided by a simple replacement of  $\tau \rightarrow -\tau$ in~\eqref{eq:math:ex5}.  Thus, the appearance of this field and its motion obey the principle of least action in the form: 
\begin{equation*}
\delta S[x^m(-\tau)]=0.
\end{equation*}
The principle of least action has two solutions: $x(\tau)$ and $x(-\tau)$. Classical mechanics considers the solution $x(\tau)$, that does not change the direction of the initial velocity, as unique. Here we have to consider both dynamic laws as a whole. This can be done using a single complex expression 
\begin{equation}\label{eq:math:ex6}
\delta \int\limits_{\tau_1}^{\tau_2} L\bigl(x(\tau),\dot{x}(\tau),\tau\bigr)\,\bigl(1-i\bigr)d\tau=0,
\end{equation}
where $\tau$ is ordinary physical time. For the principle of least action, we have 
\begin{equation*}
\delta \Bigl(S[x(\tau)]+iS[x(-\tau)]\Bigr)=0.
\end{equation*}
If the kinetic energy $T>0$ ($E>V$), then  
\begin{eqnarray}\label{eq:math:ex7}
S^+(x,\tau)=\int p\,dx - E\tau,\nonumber\\
 S^-(x,\tau)=-\int p\,dx - E\tau,
\end{eqnarray}
where $S^+(x,\tau)$ is the action field for the matter field moving along the positive direction of axis $x$, $S^-(x,\tau)$ --- in opposite direction. 
 For the momentum to become an imaginary quantity, and the kinetic energy to become negative, it is necessary to replace $\tau\rightarrow -i\tau$ in the Lagrangian in~\eqref{eq:math:ex6}. Such a substitution corresponds to the tunneling phenomenon (i.e., in this case the replacement $\tau\rightarrow -i\tau$ is not just a formal mathematical procedure). If $T<0$, then   
\begin{eqnarray}\label{eq:math:ex8}
S^+(x,\tau)=i\int p\,dx - E\tau,\nonumber\\
S^-(x,\tau)=-i\int p\,dx - E\tau,
\end{eqnarray}
where 
\begin{equation*}
p=m\frac{dx}{d\tau}
\end{equation*}
is an ordinary mechanical momentum.
Taking into account the property of additivity of the transition amplitude, for the case of transmitted and reflected waves, we have
\begin{eqnarray}\label{eq:math:ex9}
\psi(x)=A(x)\exp{\frac{i}{\hbar}S^-(x,\tau)}+B(x)\exp{\frac{i}{\hbar}S^+(x,\tau)},
\end{eqnarray}
where $A(x)$ and $B(x)$ are the functions determined by the principle of conservation of the substance.

\section{Conclusions}
It is easy to make sure that, for a rectangular potential barrier, the wave function of a tunneling particle obtained using the general expressions~\eqref{eq:math:ex9},~\eqref{eq:math:ex7} and~\eqref{eq:math:ex8} is the same as the wave functions calculated using the Schr{\"{o}}dinger equation. This coincidence necessarily follows from the fact that dynamics law~\eqref{eq:math:ex6} is directly derived from the law of quantum evolution~\eqref{eq:math:ex4}. However, the first approach makes it possible to unambiguously determine the speeds of material supports (material particles of a continuous medium) of the wave function, and the second one requires additional interpretation to do this. As mentioned above, the concept  of a corpuscle cannot underlie such an interpretation. The concept of a continuous medium as a set of matter fields uniquely interprets quantum evolution in terms of the mechanical motion of matter. Obviously, such a motion, in principle, cannot contradict special relativity, including the situation of tunneling through a potential barrier. As for the principle of locality, it is realized due to the absence of empty space and, consequently, the distance between the material particles of the considered continuous medium. In other words, the continuous medium of a quantum object is a physical reality, while the classical continuous medium is a mathematical abstraction.

The presented analysis of the passage of a quantum object through a potential barrier was carried out in order to find the material support of its quantum properties and to express the evolution of a quantum state through its mechanical motion. But the results obtained are not limited only to the interpretation of either the tunneling process or quantum effects in general. These results do not fit within the framework of orthodox quantum mechanics. In particular, the proposed approach does not contain the concept of "observable" at all, which makes it possible to use the presented method to study processes that are inaccessible for investigation by traditional methods of quantum mechanics.

And, finally, from a practical point of view, the application of the proposed approach greatly simplifies the problem of calculating the transition amplitude~\eqref{eq:math:ex2}, since it allows to avoid path integrating and directly to use expression~\eqref{eq:math:ex9}. Moreover, using it in many physical situations of interest, one can obtain an analytical expression for the transition amplitude.

\vfill\eject

\end{document}